# New approach for the incoherent and coherent combination of Gaussian laser beams and observation of the spatial thinning of the combined beams by modes phase locking.

**Hocine.Djellout, Djillali.Djellout**

Laboratoire LPCQ, UMMTO, 15000 Tizi-Ouzou
e-mail : hocine.djellout@ummto.dz

***Abstract:***

*We report a new configuration and theoretical approach for the study of the incoherent and coherent combination of Gaussian laser beams which allows us to calculate the combined intensity at every point in space, this new approach consists of directing all the lasers towards a chosen point, which is the meeting point of all the combined lasers, this method makes it possible to obtain in the case of an incoherent combination an intensity of Gaussian shape without the existence of secondary intensity lobes at the focusing plan, and in the case of a coherent combination, we observe a spatial thinning of the combined beams with a greater intensity, which depends on the numerical aperture of the system.*

## Introduction:

Since the energy and power that a single laser can provide is limited due to thermal effects and the damage threshold of the laser medium, and given the ever-increasing demand in terms of power and energy for the needs of various applications: industrial, military and fundamental research [1, 2, 3]. It is then necessary to use the technique of combining several laser beams, in order to extract more and more energy and power from the lasers. There are two approaches: the incoherent combination [4] which does not care about the phase nor the polarization and the spectral properties of the combined lasers, and the coherent combination [5-6] which requires the control of the phase of each laser beam which is therefore more difficult to achieve experimentally. The coherent combination is classified into two main categories [7-1] which are the “tiled aperture” configuration and the “filled aperture”, in the “tiled aperture” configuration the laser amplifiers are stacked side by side in a chosen geometric form, and the combination of the laser beams is done in the far field, to form a main lobe, However, there is a major problem for this configuration which limits their uses, it is the existence of secondary intensity lobes at the focusing plane of the laser beams, which alters the efficiency of combination in terms of spatial profile and intensity. To eliminate these secondary lobes, we propose a new combination configuration which consists of directing all the lasers towards a chosen point, which is the meeting point of all the combined lasers (focusing point of all the laser beams), and this either by direct propagation in free space of each beam, or by their focusing by lenses. To study this configuration, we have developed a new approach other than the Huygens Fresnel diffraction integral [8], which is used in the paraxial approximation, this model allows us to determine the intensity of the combined lasers at any point in space, in the near-field and far-field for any number of combined lasers. In our study we were interested in both the non-coherent and coherent combination of Gaussian beams, in the case of the incoherent combination, we obtain at the focal plane a combined intensity of Gaussian form, on the other hand in the case of the coherent combination, which is achieved by locking the phase of the Gaussian modes at the focusing point, we observe a spatial thinning of the combined beams with a greater intensity on the focusing axis.

## 1- Theoretical model and simulation results

In order to eliminate the secondary intensity lobes we chose a new configuration (see figure 1), which looks like the "tiled aperture", but this time, unlike the "tiled aperture" where all the beams are parallel, in the new configuration, all the beams of the source plane are oriented towards a fixed point in the observation plan. Each beam of the source plane is a Gaussian mode of waist $w_0$, and in this configuration we can have two approaches, either all the beams propagate without focusing towards the focal point of the image plane, or they are focused by lenses with focal lengths $f_{ij}$. In this study we will first treat the incoherent combination and then the coherent combination.

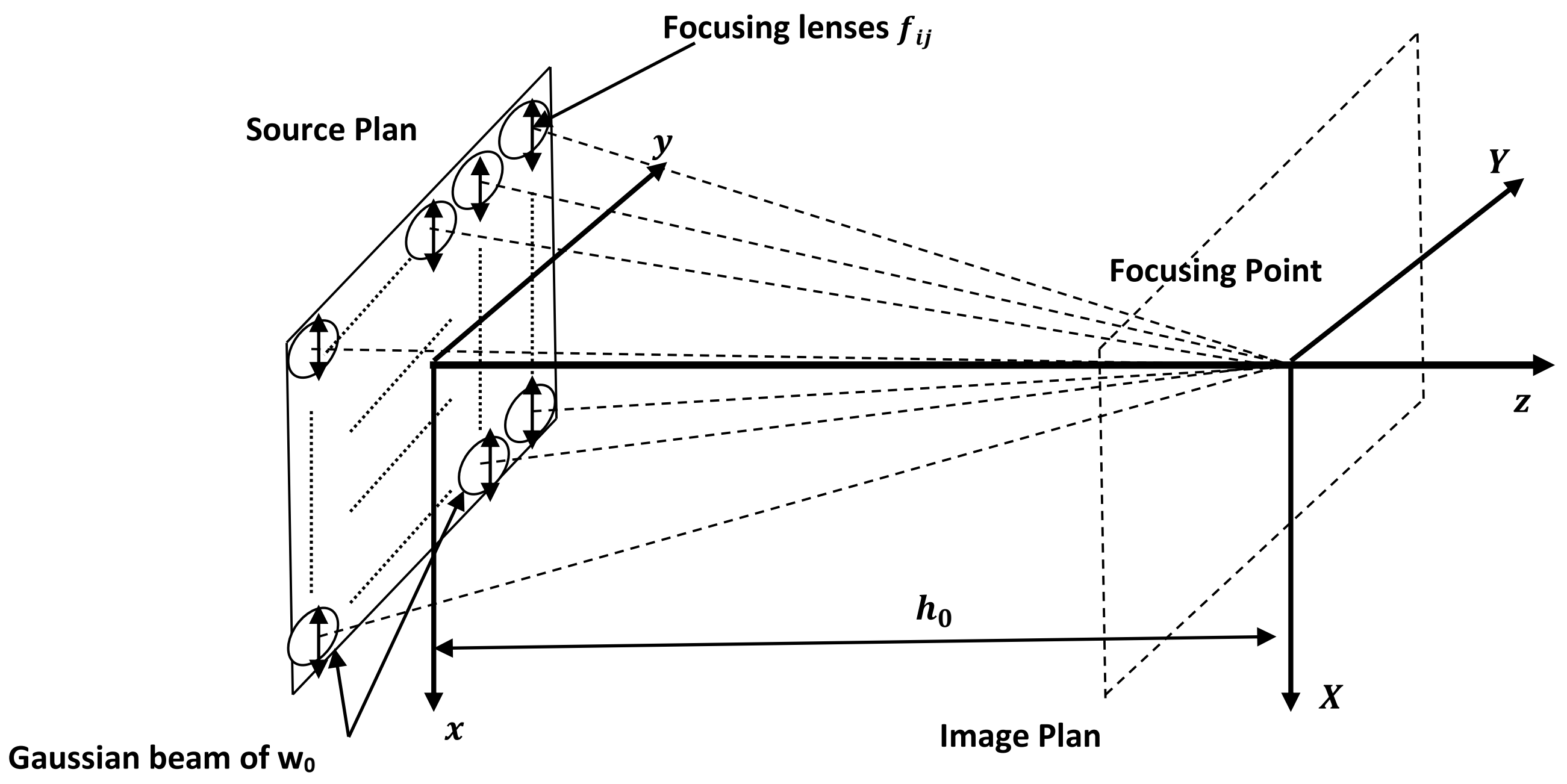

**Fig 1 : New configuration for combining Gaussian beams**

## 2.1- incoherent combination

Generally, if we know the field distribution at the source plane (at the near field), it is then possible in the paraxial approximation to determine the field at the image plane (at the far field) by the Fourier transform [9], however, in our configuration, the beams of the source plan are not parallel to each other, they are oriented towards the origin of the image plan, and their locations in the source plan are arbitrary so that the paraxial approximation is not more valid, then our idea to determine the distribution of the field at the image plane consists of the fact that since each beam of the source plane is a Gaussian mode, then we know its evolution in space, it is therefore possible to determine the distribution of its field in the image plane, and the total field distribution of all beams will be the sum of the fields of all beams in the image plane.

### 2.1.1-incoherent combination without focusing

The electric field of a Gaussian beam of a minimum waist $w_0$, whose center is at the coordinate (0,0,0) of the source plane (o,x,y,z), and which is assumed to be linearly polarized along the x axis, is described in cylindrical coordinates by this equation:

$$\overrightarrow{E_{00}} = \vec{\imath}\sqrt{I_0}\,\frac{w_0}{w(z)}\,exp\left[-\frac{r^2}{w^2(z)}\right]\,exp\left[i\left(kz - arctan\left(\frac{z}{z_0}\right) + \frac{kr^2}{2R(z)}\right)\right] \quad \textbf{(1)}$$

$I_0$ is the intensity of the beam at the origin of the z axis, $k = \frac{2\pi}{\lambda}$ , $\lambda$ is the wavelength,

$$w(z) = w_0\sqrt{1 + \left(\frac{z}{z_0}\right)^2} \quad \textbf{(2)}$$

$w(z)$ is the waist of the beam at the z coordinate, $z_0 = \frac{\pi\, w_0^2}{\lambda}$ is the Rayleigh length,

$$R(z) = z\left[1 + \left(\frac{z_0}{z}\right)^2\right] \quad \textbf{(3)}$$

$R(z)$ is the beam wavefront curvature of the Gaussian beam at the z coordinate, $r$ is the radial coordinate, and $arctan\left(\frac{z}{z_0}\right)$ is the Gouy phase shift.

The intensity of the Gaussian beam is equal to $I_{00} = \left|\left(\overrightarrow{E_{00}}\right)\left(\overrightarrow{E_{00}}\right)^*\right| = I_0\frac{{w_0}^2}{w(z)^2}\,exp\left[-2\frac{r^2}{w^2(z)}\right]$, the intensity in the image plane (O,X,Y), of equation $z = h_0$, is equal to:

$$I_{00}(\rho,\theta) = I_0\frac{{w_0}^2}{w^2(h_0)}\,exp\left[-2\frac{\rho^2}{w^2(h_0)}\right] \quad \textbf{(4)}$$

$(\rho,\theta)$ Are the polar coordinates of the image plane.

Now calculating the intensity $I_{i,j}(\rho,\theta)$ in the image plane, of a Gaussian beam whose center is located in the source plane at coordinates $(x_i, y_j)$ as shown in Figure 2.

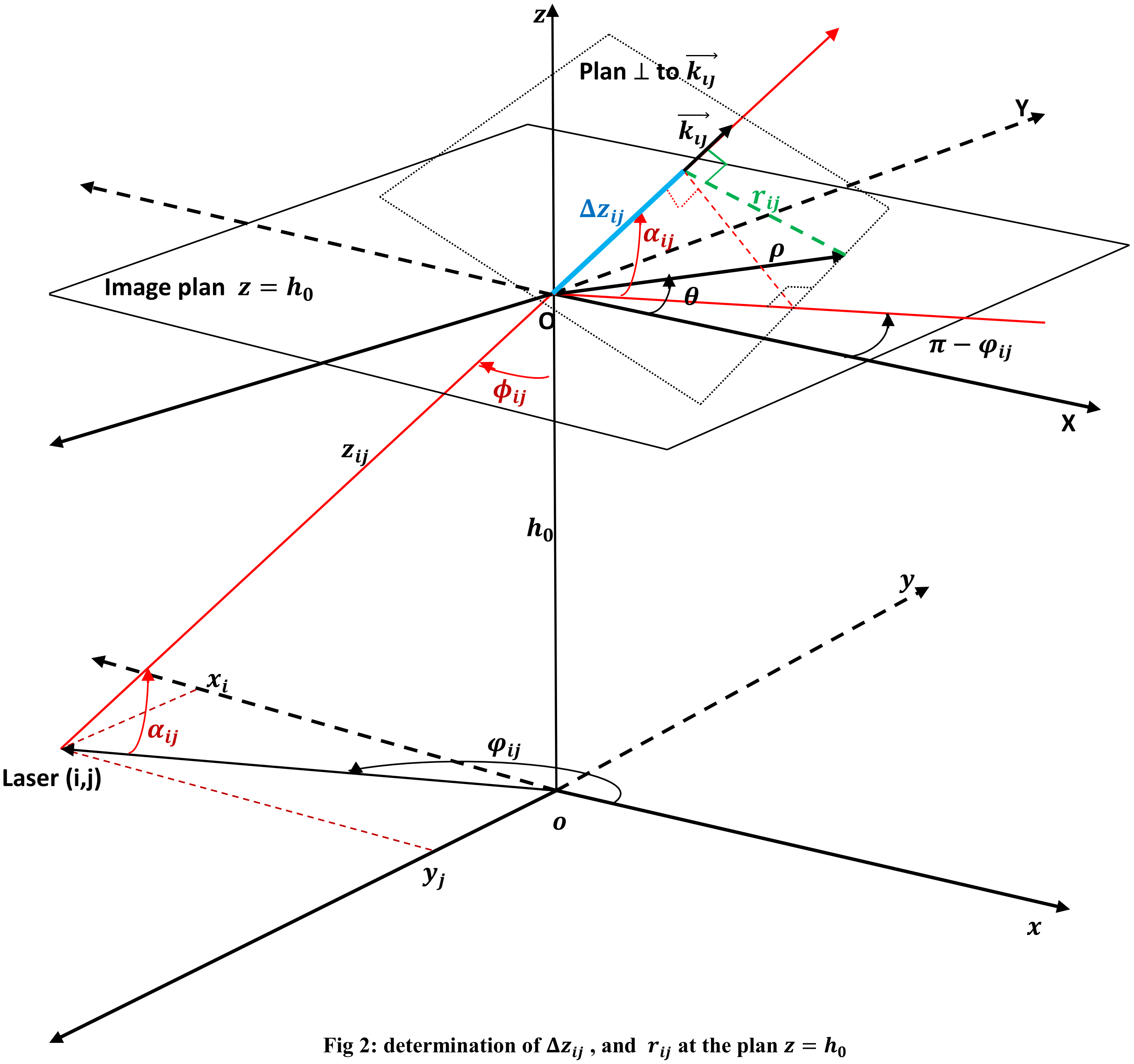

**Fig 2: determination of $\Delta z_{ij}$ , and $r_{ij}$ at the plan $z = h_0$**

The propagation of the laser beam (i,j) takes place along the $z_{ij}$ axis , parallel to the wave vector $(\overrightarrow{k_{ij}})$, it is therefore enough to know for each point of the observation planes (ρ,θ), the values of $r_{ij}$ and $\Delta z_{ij}$ corresponding to the perpendicular plan to $(\overrightarrow{k_{ij}})$, to be able to determine the value of the electric field and the optical intensity in the image plane (O,X,Y). From figure 2, we can determine $\Delta z_{ij}$ and $r_{ij}$

$$\Delta z_{ij} = \rho \cos\left(\theta - \varphi_{ij} + \pi\right) \cos\left(\alpha_{ij}\right) \quad \textbf{(5)}$$

Since $\Delta z_{ij}$ , $r_{ij}$ and ρ forms a right triangle, then $r_{ij}$ is equal to

$$r_{ij} = \rho\sqrt{1 - cos^2\left(\theta - \varphi_{ij} + \pi\right) cos^2\left(\alpha_{ij}\right)} \quad \textbf{(6)}$$

Thus the laser intensity (i,j) in the image plane is equal

$$I_{i,j}(\rho,\theta) = I_0 \frac{{w_0}^2}{w^2(z_{ij}+\Delta z_{ij})}\, exp\left[-2\frac{\rho^2\left(1-cos^2(\theta-\varphi_{ij}+\pi)\, cos^2(\alpha_{ij})\right)}{w^2(z_{ij}+\Delta z_{ij})}\right] \quad \textbf{(7)}$$

$z_{ij} = \sqrt{{x_i}^2 + {y_j}^2 + {h_0}^2}$ Is the distance between the position of the laser (i,j) in the source plane and the origin of the image plane, $\varphi_{ij}$ is the polar angle identifying the laser (i,j) in the source plane (o,x,y).

Since in this case, the combined laser beams are incoherent, then the total intensity $I_T(\rho,\theta)$ that we measure on the image plane will be the sum over all the laser intensities.

$$I_T(\rho,\theta) = \sum_{i,j} I_{i,j}(\rho,\theta) \quad \textbf{(8)}$$

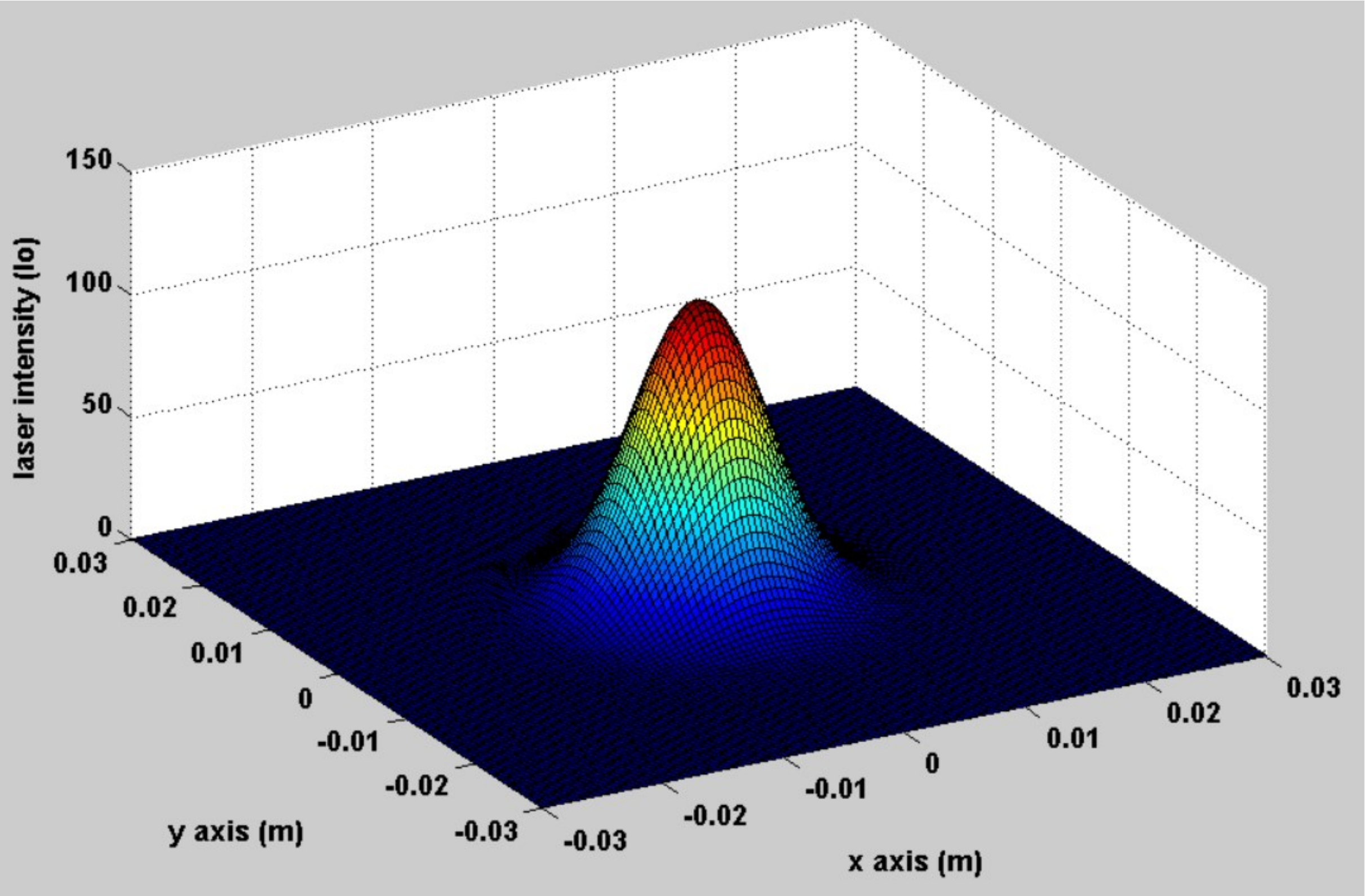

**Fig 3:** combined intensity of 121 incoherent laser beams at the focal point ($h_0 = 1m$)

Figure 3 shows us the result of combinations without focusing with lenses of 121 incoherent Gaussian laser beams with a minimum waist of 0.01 m, wavelength of 1 μm, assembled at the plan source with a square shape. The intensity obtained is a Gaussian shape with a beam radius (waist) of approximately 0.01 m and an intensity at the axis of 121 $I_0$. Indeed, since the Rayleigh length $z_0 = 314\, m$, therefore the Gaussian beam does not change over a distance of $h_0 = 1m$, which explains that the intensity obtained at the focal point is equal to the sum of all the intensities of the laser beams.

After having calculated the intensity at the image plane of equation $z = h_0$, let us now calculate the total intensity $I_T(X, Y, h)$ at any plan of equation $z = h$, where $h$ is the distance between the latter and the source plane, as shown in Figure 4.

Let O' be the point of intersection between the plan of equation $z = h$ and the line $z_{ij}$, the coordinates of O' are $(x_{0i}, y_{0j}, h)$ as shown in figure 4, such that $x_{0i} = x_i\left(1 - \frac{h}{h_0}\right)$ and $y_{0j} = y_j\left(1 - \frac{h}{h_0}\right)$. Now considering (O',X',Y') the new reference frame in which we will calculate the distribution of the optical intensity of only the laser (i,j) so that we obtain equations similar to ( 5) and (6), thus we obtain

$$\Delta z'_{i,j} = \rho' \cos(\theta' - \varphi_{ij} + \pi) \cos(\alpha_{ij}) \qquad \textbf{(9)}$$

$$r'_{i,j} = \rho' \sqrt{1 - cos^2(\theta' - \varphi_{ij} + \pi)\, cos^2(\alpha_{ij})} \qquad \textbf{(10)}$$

But since our goal is to calculate the total intensity of all the laser beams at the plan $z = h$, and in the reference frame (O,X,Y), therefore we must express the coordinates (ρ',θ') as a function of X and Y. as $\overrightarrow{OM} = \overrightarrow{OO'} + \overrightarrow{O'M}$ as shown in Figure 4, then $X' = X - x_{0i}$ and $Y' = Y - y_{0j}$ so that $\rho' = \sqrt{(X - x_{0i})^2 + (Y - y_{0j})^2}$ and $\theta' = arctang\left(\frac{Y - y_{0j}}{X - x_{0i}}\right)$. The intensity of the laser (i,j) at the plan $z = h$ is then equal to

$$I_{i,j}(X, Y, h) = I_0 \frac{w_0{}^2}{w^2(z'_{i,j} + \Delta z'_{i,j})}\, exp\left[-2 \frac{\rho'^2\left(1 - cos^2(\theta' - \varphi_{ij} + \pi)\, cos^2(\alpha_{ij})\right)}{w^2(z'_{i,j} + \Delta z'_{i,j})}\right] \qquad \textbf{(11)}$$

$z'_{i,j} = \frac{h}{h_0} z_{ij}$ Is the distance between the position of the laser (i,j) in the source plane and the point $O'$.

The total intensity $I_T(X, Y, h)$ that we measure in the plan $z = h$ is the sum of the intensities of all the lasers in this plan.

$$I_T(X, Y, h) = \sum_{i,j} I_{i,j}(X, Y, h) \qquad \textbf{(12)}$$

Note that we find back the intensity of equation (8) for $h = h_0$

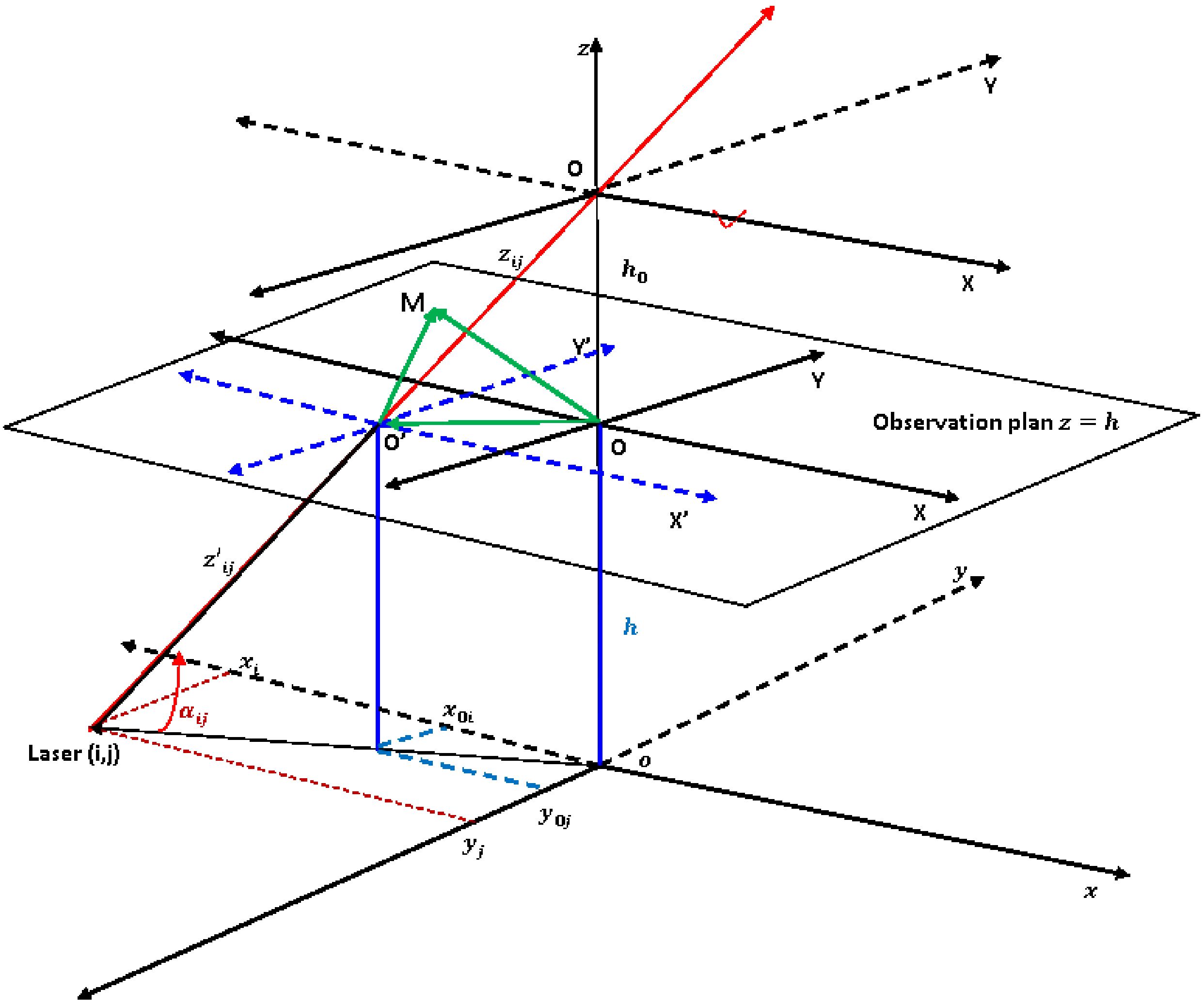

**Fig 4: determination of $\Delta z_{ij}{}'$, and $r_{ij}{}'$ at the plan $z = h$**

### 2.1.2- incoherent combination with focusing

The previous study is carried out for an incoherent combination of Gaussian beams propagating in free space, in this part we are interested in their combinations by focusing them with lenses, so that each Gaussian beam is focused at the focal point which is the origin of the image plane as shown in Figure 1. For this we need to know how a Gaussian beam transforms after passing through a lens, consider a Gaussian beam having a minimum waist $w_0$ at the level of the lens of focal length $f$ as shown in figure 5. By using the ABCD matrices and the complex beam parameter, we find:

$$\begin{pmatrix} A & B \\ C & D \end{pmatrix} = \begin{pmatrix} 1 & z \\ 0 & 1 \end{pmatrix} \begin{pmatrix} 1 & 0 \\ \frac{-1}{f} & 1 \end{pmatrix} = \begin{pmatrix} 1 - \frac{z}{f} & z \\ \frac{-1}{f} & 1 \end{pmatrix} \qquad \textbf{(13)}$$

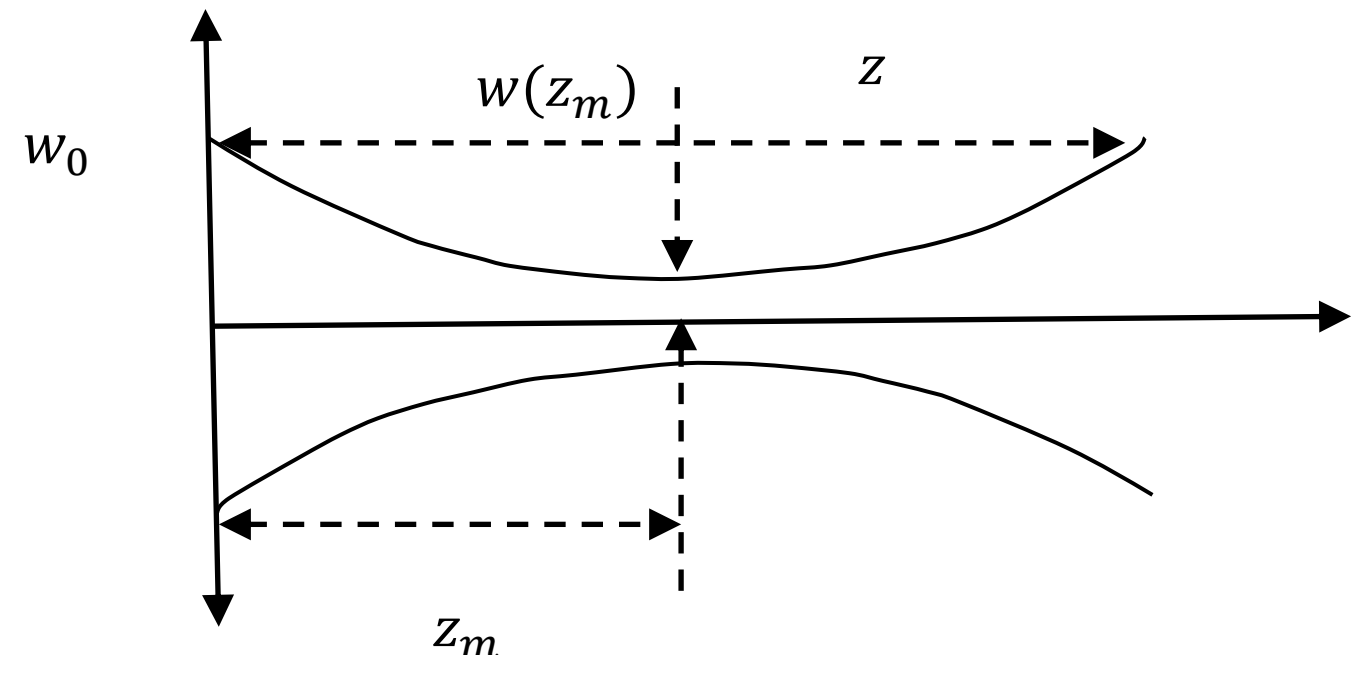

**Fig 5:** focusing of a Gaussian laser beam by a lens

After calculating we find

$$R(z) = \frac{\left(1-\frac{z}{f}\right)^2+\left(\frac{z}{z_0}\right)^2}{\frac{-1}{f}+z\left(\frac{1}{f^2}+\frac{1}{z_0^2}\right)} \qquad \textbf{(14)}$$

$$w(z) = \sqrt{\frac{\lambda}{\pi}}\sqrt{\frac{\left(1-\frac{z}{f}\right)^2+\left(\frac{z}{z_0}\right)^2}{\frac{1}{z_0}}} \qquad \textbf{(15)}$$

$R(z)$ and $w(z)$ are respectively the beam curvature and the beam waist at the z coordinate, $z_m$ is the position of the new minimum waist of the laser beam after passing through the lens, we can find this position from equation (14) with the condition $1/R(z_m) = 0$ , after calculating we find

$$z_m = \frac{f}{1+\left(\frac{f}{z_0}\right)^2} \qquad \textbf{(16)}$$

In our case the Rayleigh length is much greater than the focal length of the lens $z_0 \gg f$ , then, we can say that the minimum waist is approximately at the focus of the lens $z_m \approx f$.

Since the distance separating the laser location in the source plane from the focal point is different for each laser (see figure 1), then, to have a better combination efficiency, it is necessary to focus each laser with a lens of focal length $f_{ij} = z_{ij}$, moreover since we always have the same Gaussian symmetry for the focused beams, we just need to replace $z$ by $z'_{i,j} + \Delta z'_{i,j}$ in equation (15) and thus be able to calculate the intensity distribution at the image plane of equation $z = h$ by replacing the focused $w(z)$ of equation (17) in equation (11).

$$w\left(z'_{i,j} + \Delta z'_{i,j}\right) = \sqrt{\frac{\lambda}{\pi}}\sqrt{\frac{\left(1-\frac{z'_{i,j}+\Delta z'_{i,j}}{z_{ij}}\right)^2+\left(\frac{z'_{i,j}+\Delta z'_{i,j}}{z_0}\right)^2}{\frac{1}{z_0}}} \qquad \textbf{(17)}$$

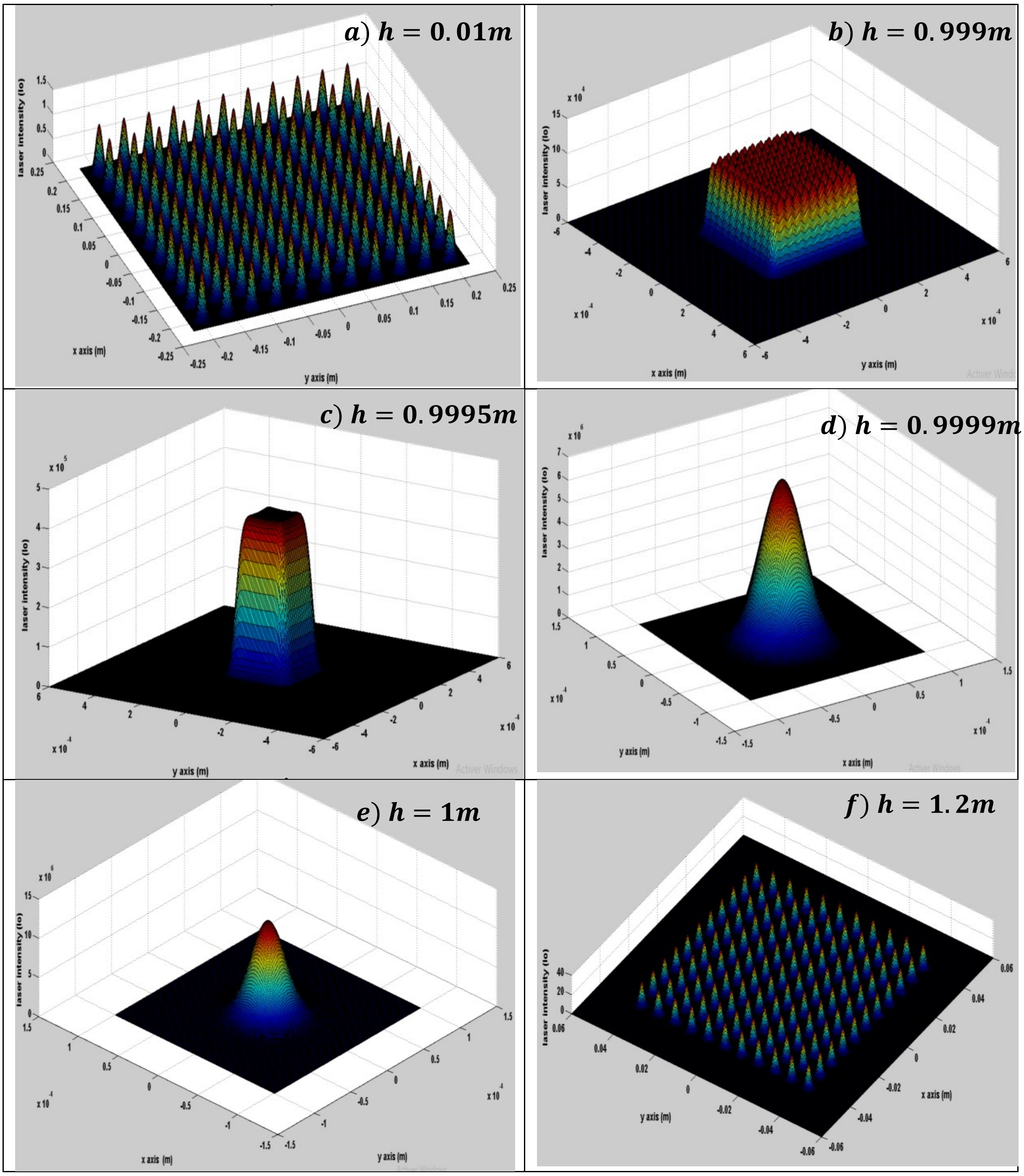

**Fig 6: spatial evolution of the combined intensity of 121 focused incoherent lasers**

Figure 6 shows us the spatial evolution of the combined intensity of 121 incoherent Gaussian beams, focused by lenses at the focal point which is at a distance $h_0 = 1\ m$ from the source plane, Figure 6 (a) shows us that for a distance $h = 0.01m$ the focused beams do not overlap, and they keep the same Gaussian shape. However, from around $h = 0.999m$ (figure 6 (b)) we see that there is partial overlap between the beams, and they form a square wavefront with small bumps on the surface, the combined intensity is 1.115 $10^5$ $(Io)$ focused into an area of 208µmX208µm. at a distance $h = 0.9995\ m$ (figure 6 (c)) the combined beams have a square-shaped wavefront without bumps with an area of 100µmX100µm and an intensity of 3.951 $10^5(Io)$. From $h = 0.9999m$ (figure 6 (d)), the combined intensity has a Gaussian shape with a waist of 40 µm and an intensity of 6.903 $10^6(Io)$. At the focal point for $h = h_0 = 1\ m$, there is total overlap between the beams which form a Gaussian with a waist of approximately 33 µm (figure 6 (e)), and an intensity of 1.158 $10^7\ (Io)$. For $h = 1.2\ m$ (figure 6 (f)), again, there is no overlap between the beams and disperse more quickly after their focusing at the focal point.

### 2.2- coherent combination

Now we will be interested in the coherent combination of Gaussian beams with always the same configuration of Figure 1, in this case, unlike the incoherent combination, the phase and polarization of each laser beam is important, because the goal is to put all the laser beams at the same phase at the focal point in the image plane, so that the electric field of all lasers interferes in a constructive way, for this all lasers must have the same polarization state, for example along the x axis. However, since in our configuration, each laser in the source plane must be oriented towards the focal point, and since the electric field is perpendicular to the direction of propagation, then the electric field cannot be only in the x direction. The equation of the plan in the focal point which is perpendicular to the direction of propagation of the laser (i,j) is given by $\overrightarrow{OM}\ \frac{\overrightarrow{k_{ij}}}{\|\overrightarrow{k_{ij}}\|} = 0$

$$-x_i X - y_j Y + h_0 z = 0 \qquad \textbf{(18)}$$

From equation (18), we notice that by rotating the laser along its propagation axis, that it is possible to obtain two directions for the polarization, so that the amplitude of the electric field is writing:

$$\overrightarrow{E_{0ij}} = E_0 \left( \frac{h_0}{\sqrt{x_i^2 + h_0^2}} \vec{\imath} + \frac{x_i}{\sqrt{x_i^2 + h_0^2}} \overrightarrow{e_z} \right) \qquad \textbf{(19)}$$

Thus the electric field at the focal point produced by a Gaussian beam of minimum waist $w_0$, whose center is at the coordinate (i,j) of the source plane (o,x,y,z) is described by this equation:

$$\overrightarrow{E_{ij}}(\rho,\theta,h_0) = \sqrt{I_0} \left( \frac{h_0}{\sqrt{x_i^2 + h_0^2}} \vec{\imath} + \frac{x_i}{\sqrt{x_i^2 + h_0^2}} \overrightarrow{e_z} \right) \frac{w_0}{w(z_{ij} + \Delta z_{ij})}\ exp\left[ -\frac{{r_{ij}}^2}{w^2(z_{ij} + \Delta z_{ij})} \right]\ exp\left[ i\left( k(z_{ij} + \Delta z_{ij}) - arctan\left( \frac{z_{ij} + \Delta z_{ij}}{z_0} \right) + \frac{k{r_{ij}}^2}{2R(z_{ij} + \Delta z_{ij})} \right) \right] \qquad \textbf{(20)}$$

In the case where the laser beam (i,j) is not focused, then we must use the expression of $w(z)$ of equation (2), on the other hand, if the laser is focused, then, in this case we must use the expression of $w(z)$ of equation (15), moreover, if we want to calculate the distribution of the intensity at the image plane of equation $z = h$, then we must use the expressions of $\Delta z_{i,j}^{'}$ and $r_{i,j}^{'}$ of equations (9) and (10).

As the laser beam at the focal point is practically a localized plane wave, then the beam curvature R(z) tends towards infinity, which implies that $\frac{k{r_{ij}}^2}{2R(z_{ij} + \Delta z_{ij})}$ tends to zero, similarly $z_{ij} \ll z_0$ so that we can neglect the term $arctan\left( \frac{z_{ij} + \Delta z_{ij}}{z_0} \right)$.

To have constructive interference between all beams at the focal point, it is necessary that the phase of each laser is locked at the focal point, such that $k\, z_{ij} - \omega t_{ij} = 0$ for all lasers (i,j), $t_{ij} = \frac{z_{ij}}{C}$, to achieve this, it is therefore necessary to use in the experimental device an optical path difference controller, and a phase locking system [7].

With these approximations, and by replacing the term $\cos(\alpha_{ij})$ by $\sin(\phi_{ij})$ where $\phi_{ij}$ is the angle at which we see the laser (i,j) from the image focal point (see figure 2), such that $\sin(\phi_{ij})$ represents the numerical aperture under which we see the laser (i,j). Thus the total electric field produced by all the lasers (i,j) at the focal point is written:

$$\overrightarrow{E_T}(\rho,\theta,h_0) = \sum_{i,j}\sqrt{I_0}\left(\frac{h_0}{\sqrt{x_i^2+h_0^2}}\vec{i} + \frac{x_i}{\sqrt{x_i^2+h_0^2}}\overrightarrow{e_z}\right)\frac{w_0}{w(z_{ij}+\Delta z_{ij})}\, exp\left[-\frac{{r_{ij}}^2}{w^2(z_{ij}+\Delta z_{ij})}\right] exp\left(i2\pi\,\rho\,\cos(\theta-\varphi_{ij}+\pi)\,\frac{\sin(\phi_{ij})}{\lambda}\right) \quad \textbf{(21)}$$

The total intensity at the image plane is then equal

$$I_T(\rho,\theta,h_0) = \left|\overrightarrow{E_T}(\rho,\theta,h_0)\overrightarrow{E_T}^*(\rho,\theta,h_0)\right| \quad \textbf{(22)}$$

Figure 7 shows us the spatial evolution of the combined intensity of 121 coherent Gaussian beams focused by lenses, Figure 7 (a) shows us that for a distance from the image plan of $h = 0.01m$, the focused beams do not overlap, and we therefore observe no interference between the beams, and thus we obtain the same thing as in the incoherent case (see figure 6(a)). For $h$=0.999$m$ (figure 7 (b)) we see that there is partial overlap between the beams, and we observe partials interferences between the beams, because we obtain 196 peaks with secondary lobes, instead of 121, these peaks are focused on a square surface of 440µmX440µm, where each peak has an intensity of 4.231 $10^5(Io)$, moreover the comparison between the incoherent case (figure 6 (b)) and the coherent case (figure 7 (b)), shows us different results. For $h = 0.9995\ m$ (figure 7 (c)), the focusing surface decreased to 226µmX226µm and the intensity of the peaks increased to 1.248 $10^6(Io)$. On the other hand, at $h = 0.9999\ m$ (figure 7 (d)), we observe 4 great peaks of intensity 5.604 $10^7(Io)$ surrounded by a few peaks of weaker intensity, however, in the case of the incoherent combination, we obtained a Gaussian shape see (figure 6 (d)), from which we understand that the coherent combination requires more precision than the coherent combination. For $h = 1.2m$ (figure 7 (f)) we obtain practically the same thing with the incoherent case, because there is no overlap between the beams, and they disperse more quickly after their focusing at the focal point.

At the focal point $h = h_0 = 1\ m$, we observe a main peak of Gaussian shape surrounded by secondary lobes of lower intensity which align along the x and y axes which have the shape of a cardinal sinus (figure 7 (e)), which are the Fourier transform of a two-dimensional square function, because in fact, the combined laser beams are assembled at the source plane with a square shape. The main peak has a Gaussian shape with a waist of approximately 1.65 µm, and an intensity of 1.379 $10^9(Io)$ which is very slightly lower than 121 times the intensity of the non-coherent case 1.158 $10^7$ $(Io)$ which can be explained by the existence of a slight polarization in the direction of the z axis, which is due to the fact that the beams are slightly tilted towards the focal point. In addition, we note a thinning of the combined beams at the focal plane (a waist of 1.65 µm), compared to the incoherent combination (waist of 33 µm), note that we observe the same thinning (a waist of 1.65µm) as shown in Figure 8 for the coherent combination of 121 lasers in free space but without focusing them with lenses. We notice here an analogy with temporal mode locking which results from the summation of the modes in the frequency domain, which shortens the temporal width of the pulse, while in our case, it is a summation of the Gaussian modes in the angular domain, which results from the term $\sin(\phi_{ij})$ of equation (21) which causes the focused intensity to spatially thin. Indeed, we can explain the limit of diffraction by the principle of uncertainty which is due to the Fourier transform, where we can write

$$\Delta\{\rho\ \cos(\theta - \varphi_{ij} + \pi)\}\Delta\left\{\frac{\sin(\phi_{ij})}{\lambda}\right\} \geq \frac{1}{2} \quad \textbf{(22)}$$

By resonating in a single dimension, so that we set $\cos(\theta - \varphi_{ij} + \pi) = 1$, then we obtain

$$\Delta\rho \geq \frac{\lambda}{2\Delta\{\sin(\phi_{ij})\}} = \frac{\lambda}{2\,ON} \quad \textbf{(23)}$$

Equation (23) shows us that if we have an incertainty on the angle from which the photons come of $\sin(\phi_{ij}) = ON$ , then we have an incertainty on the position in the image plan of $\Delta\rho = \lambda\ /(2ON)$. The maximum incertainty on the angle that can be obtained is $\phi_{ij} = \pi/2$ which corresponds to a numerical aperture of $ON = 1$, to the latter corresponds an incertainty of $\Delta\rho = \lambda/2$. Conversely, if we know precisely the angle from which the photons come $\phi_{ij} = 0$ , then we have complete incertainty (infinity) on the position in the image plane.

We can also find the angular diffraction deviation which is given by $\theta = \lambda/a$ from the incertainty relation $\Delta\rho\ ON\ \geq \lambda/2$, where $(a)$ represents the diameter of the aperture in the source plan. As $ON = a/(2\ h)$, where $h$ is the distance between the source and image plan, and as $\theta \approx \Delta\rho/h$ then we obtain $\theta \geq \lambda/a$.

by increasing the surface of the plane source, we thus increase the numerical aperture of the system which results in a reduction in the focal spot of the combined intensity as shown in Figure 9, which is obtained by increasing the surface of the source plane to 4 m2, with just over 5000 combined lasers, the waist of the focused intensity is less than the wavelength (0.5 µm). To further thin the focused intensity, it is necessary to increase the numerical aperture, and for this, the planar geometry of the source plane is not effective, it is more judicious to use a curved source plane, such as for example a half sphere, thus increasing the numerical aperture to a value of 1. On the other hand, if we want to obtain greater intensity, planar geometry is more interesting because it can contain more laser beams.

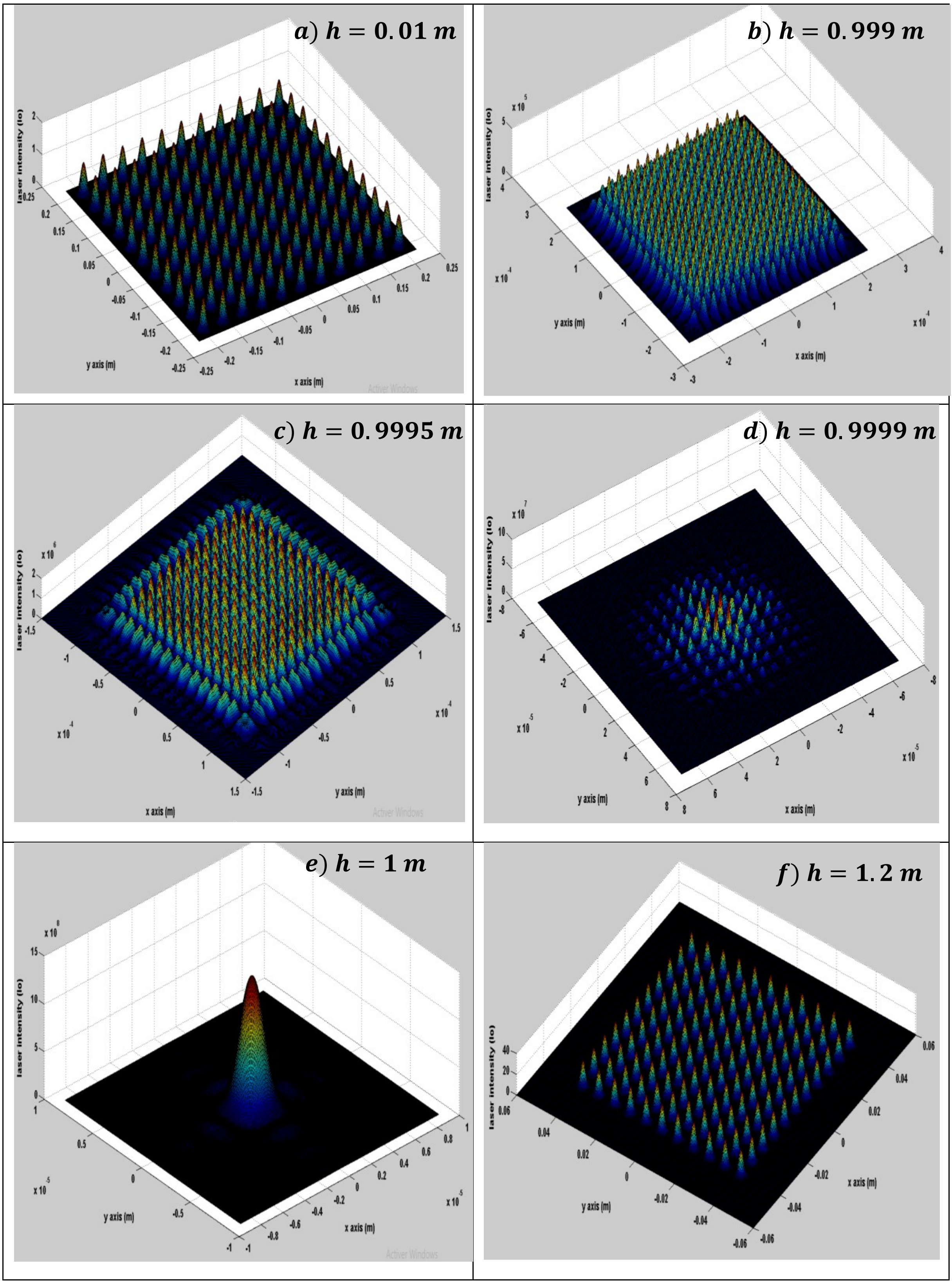

**Fig 7: Spatial evolution of the combined intensity of 121 focused coherent lasers**

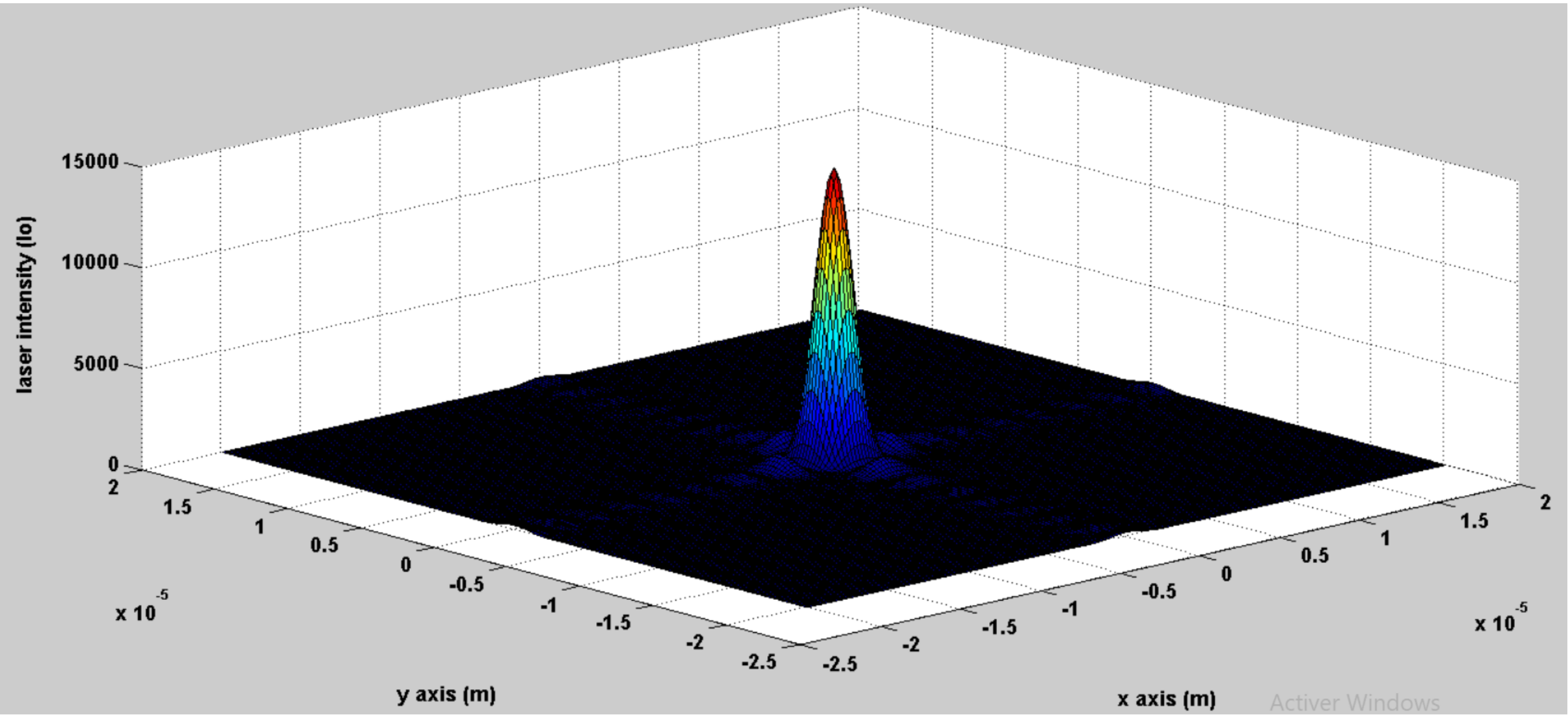

**Fig 8 : Combined intensity of 121 coherent lasers not focused by lenses**

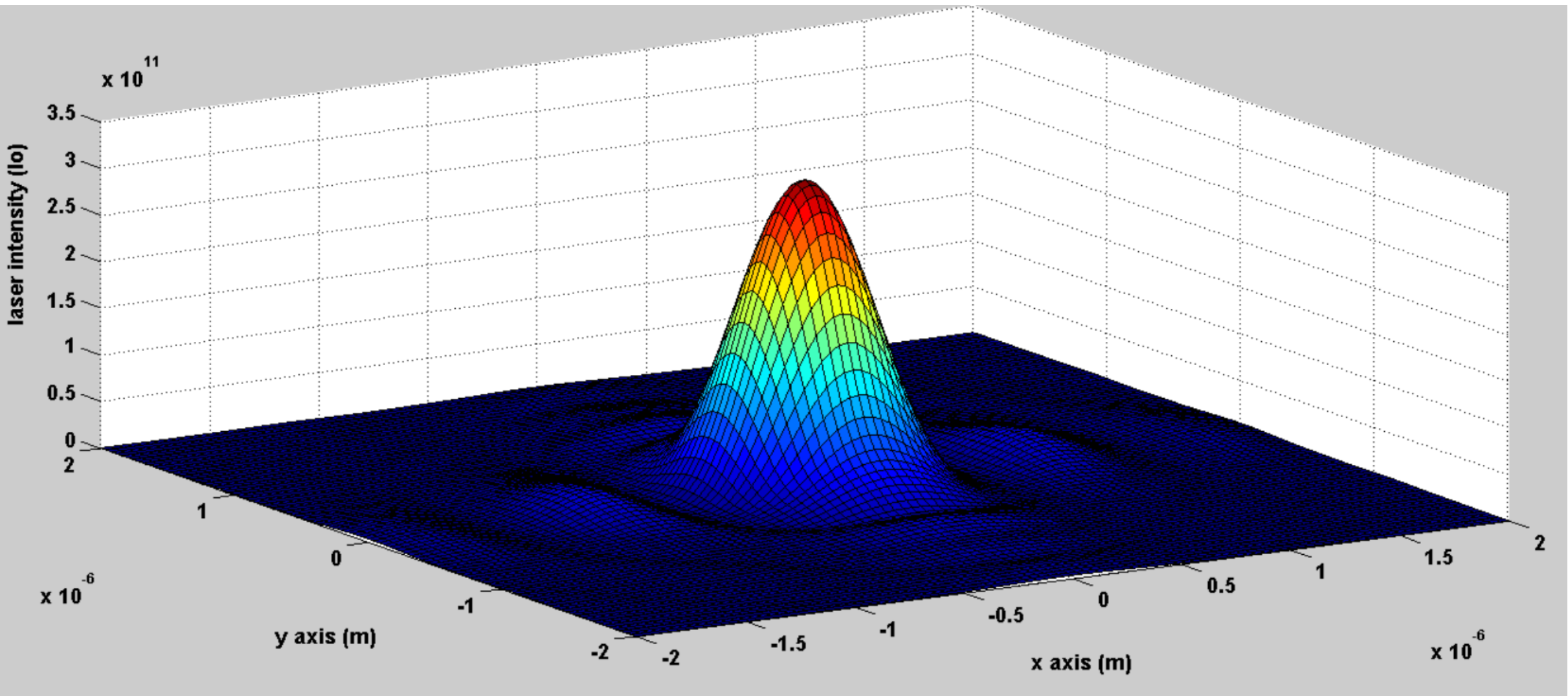

**Fig 9 : Combined intensity of more than 5000 focused coherent lasers, the source plane has an area of 4 m2**

The intensity obtained with focusing the beams is much greater than in the case without focusing because all the intensity of the beams are focused with lenses at the point $z = h_0$, moreover, in the case of the coherent combination, the obtained intensity is approximately N times the intensity of the non-coherent case, where N is the number of combined beams. by increasing the numerical aperture, namely, by focusing this time at a distance of 0.25 m ($h = h_0 = 0.25m$) instead of 1 m, we obtain from the combinations of 121 laser beams, an intensity of 1.304 $10^{10}$ (Io) with a waist of 0.52 µm, if the power of each combined laser is 6 KW in continuous mode (CW), we obtain an intensity in the axis of approximately 7 $10^{13}$ W/cm2 in continuous mode , if we use pulsed lasers of 6 GW and 200 fs [10], we obtain an intensity in the axis of approximately 7 $10^{19}$ W/cm2, with such an intensity the matter instantly becomes a plasma, and the movement of Electrons become relativistic, and protons can be accelerated. Noting that in the impulsionel case, to have exact calculations, it is necessary to take into account the temporal evolution of the beams, however, we believe that the intensity in the axis will be practically the same as in the continuous case, because the beams arrive at the focal point at the same time, moreover, since the waist of the combined intensity is a few µm, which is also of the same order of magnitude as $\Delta z_{ij}$, which will correspond to advances or delays at the focal plane of a few tens of femtosecond (fs), then, for large pulses greater than a few hundred fs, we expect to find practically the same spatial distribution as in the continuous case.

We believe that with this new configuration it is possible to reach record in intensities, because we are no more limited by power management problems since all the laser beams propagate in free space.

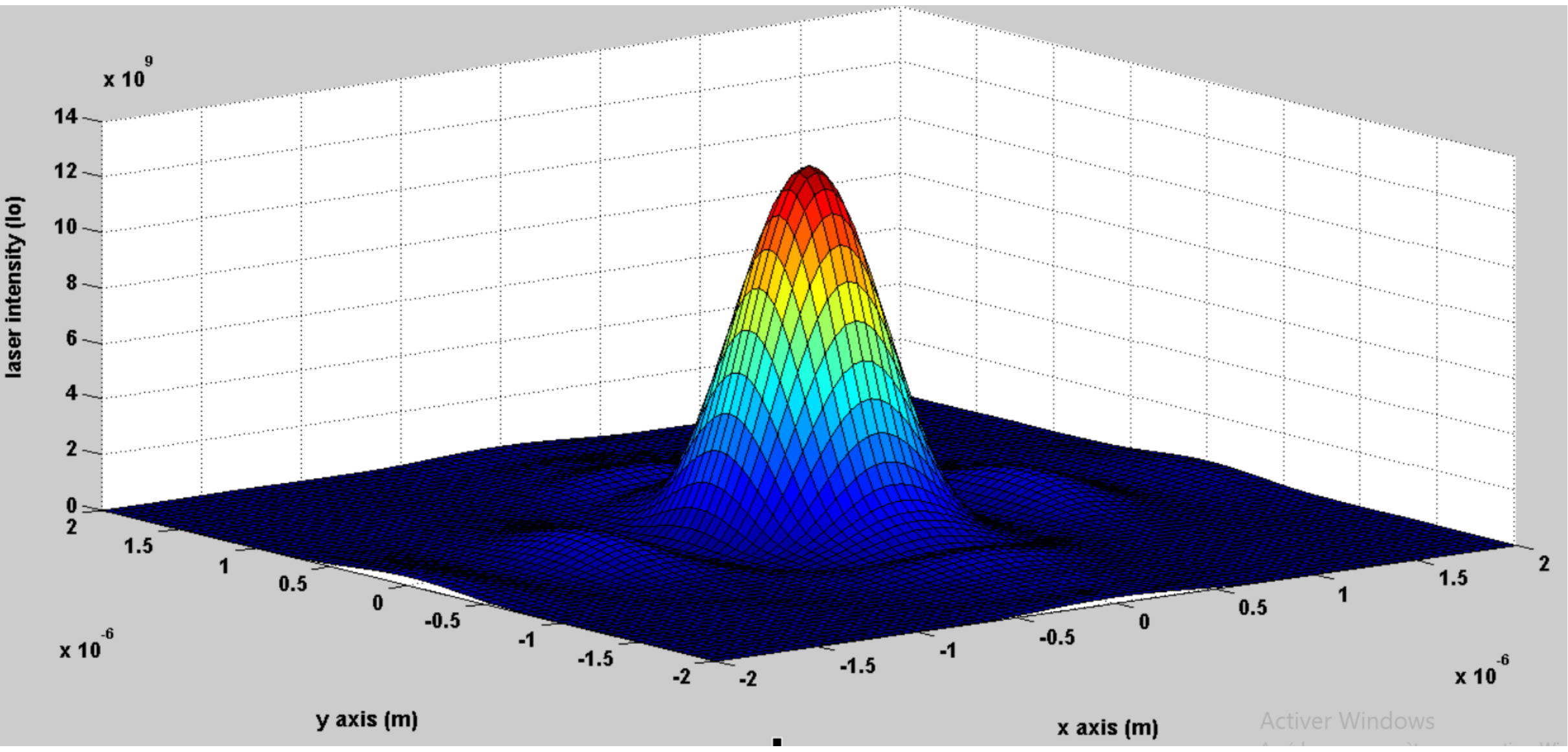

**Fig 10 : Combined intensity of 121 coherent lasers focused at 25 cm**

Contrary to what is reported in the literature [1,7,11], we find in our simulations that both in the coherent and incoherent combination, that the filling factor (the minimum gap between individual laser beams) has no impact on the formation of the secondary lobes as well as on the proportion of energy that the central lobe takes, in fact, in the tiled aperture configuration, the laser beams are parallel to each other in the plane source, and the combination takes place in the far field, but since the beams are initially parallel, and the maximum intensity of each beam is always on its propagation axis, then, there cannot be a total combination of laser beams even in the far field, but there will be have a partial combination between the beams, and this explains the formation of the side lobes. It is then essential to direct all the laser beams towards the same point (focusing point) with very high precision to have a total combination of the laser beams and thus eliminate the formation of secondary lobes.

## Conclusion

In this work, we proposed a new configuration for the coherent and incoherent combination of Gaussian laser beams, this configuration makes it possible to efficiently combine the lasers, and eliminates the existence of side lobes. To study this configuration a new theoretical approach is developed, which allows us to determine the field and the total combined intensity at all points in space. In the case of a coherent combination, we observe a spatial thinning of the combined intensity, which is due to the increase in the numerical aperture of the system.

**Références**